Figure 1

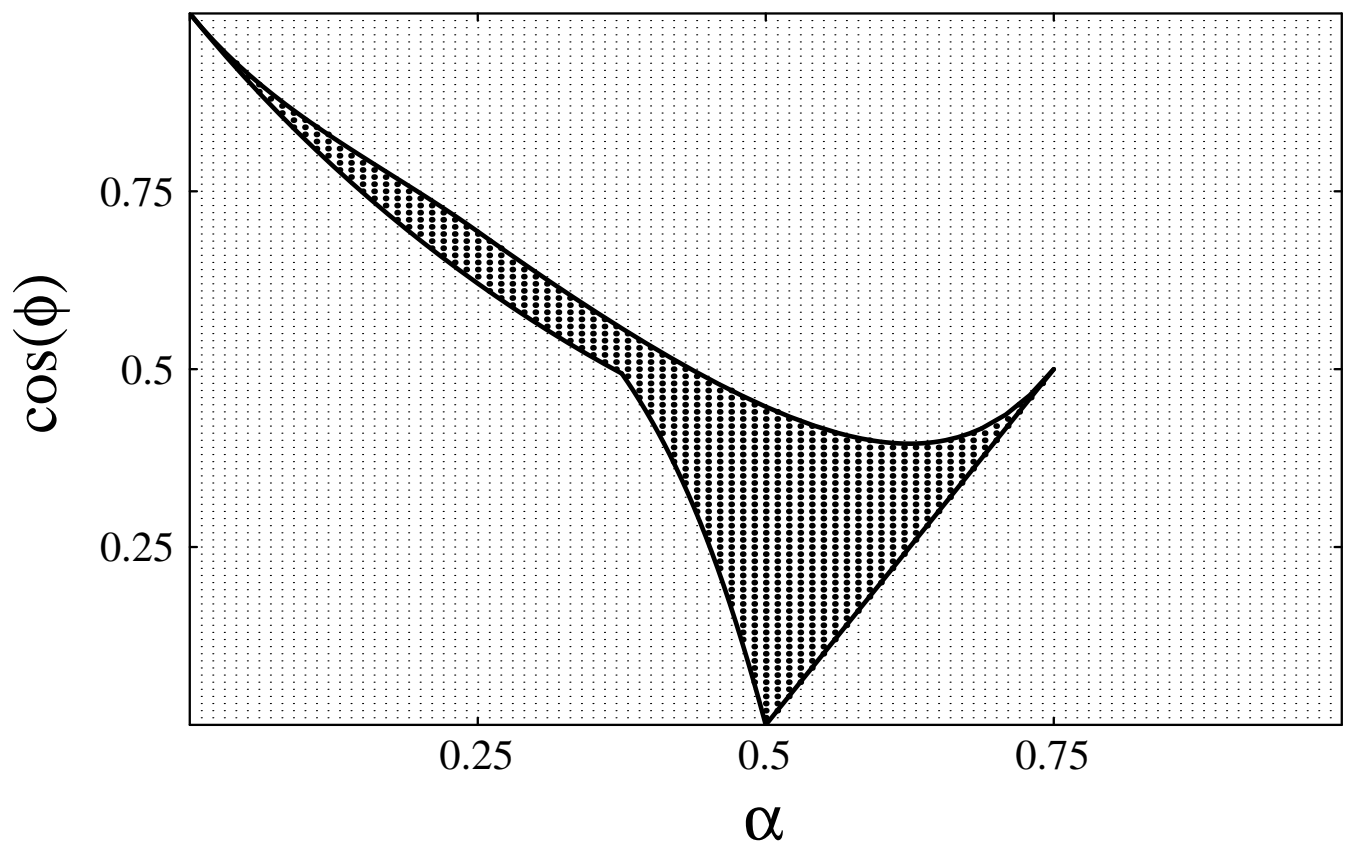

Figure 2a

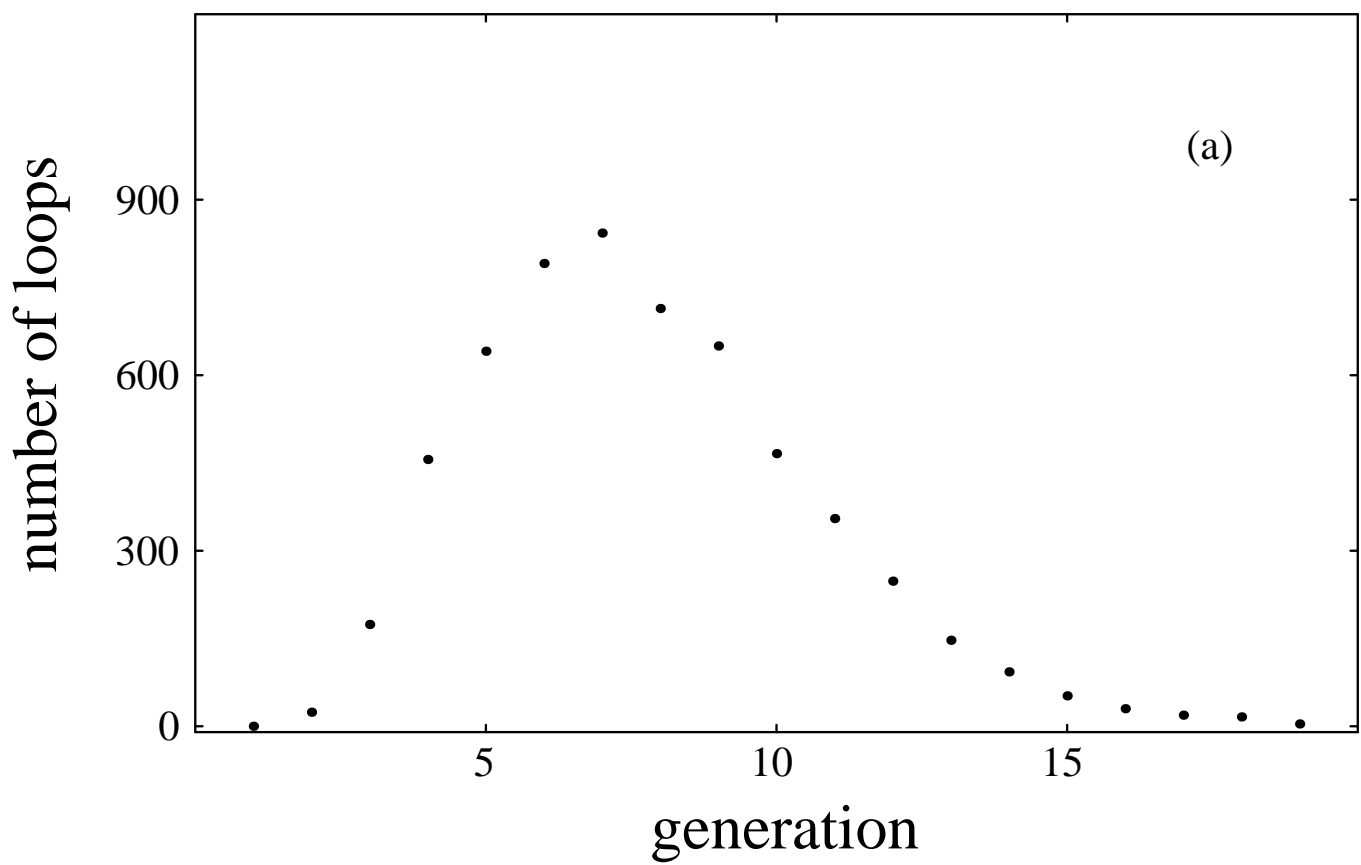

Figure 2b

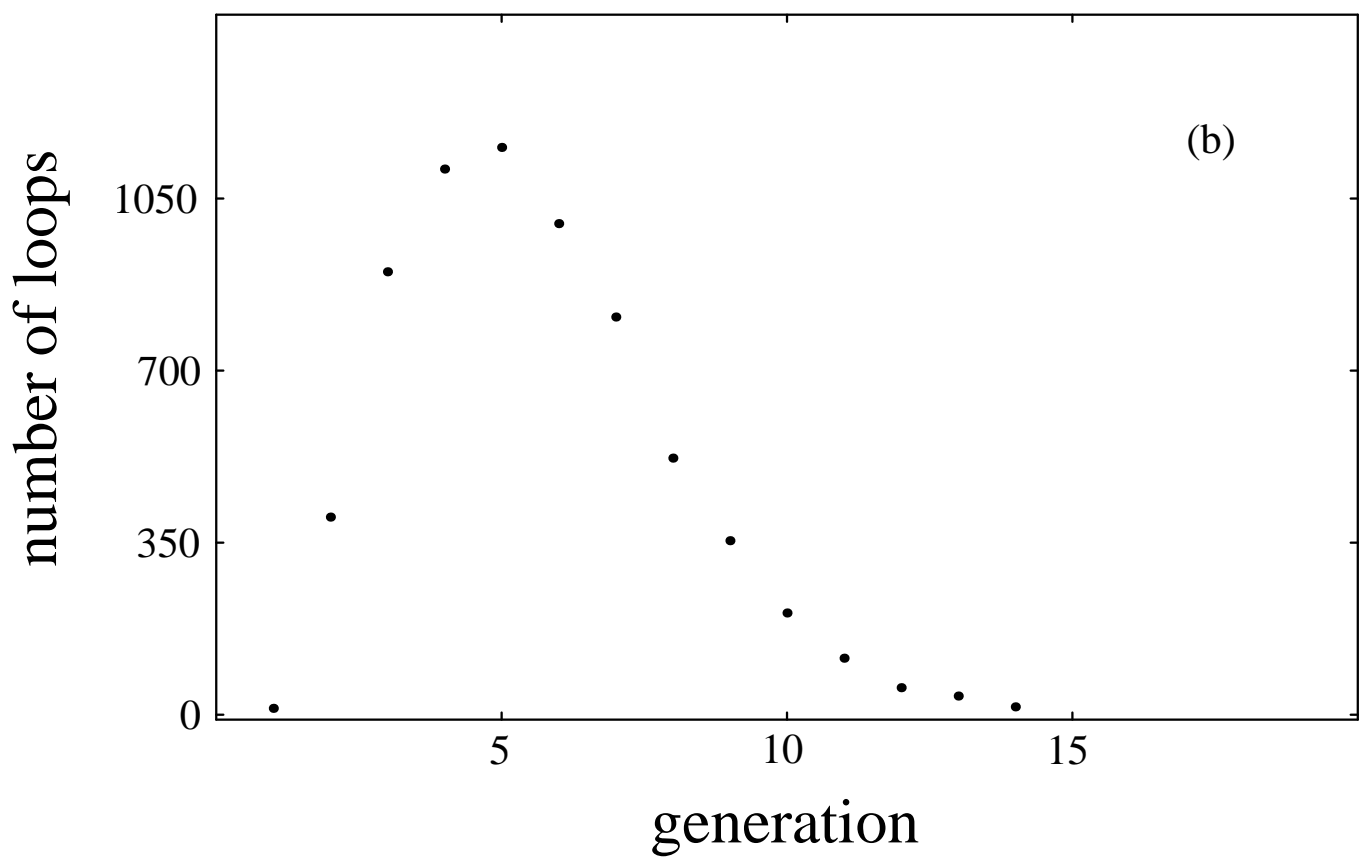

Figure 2c

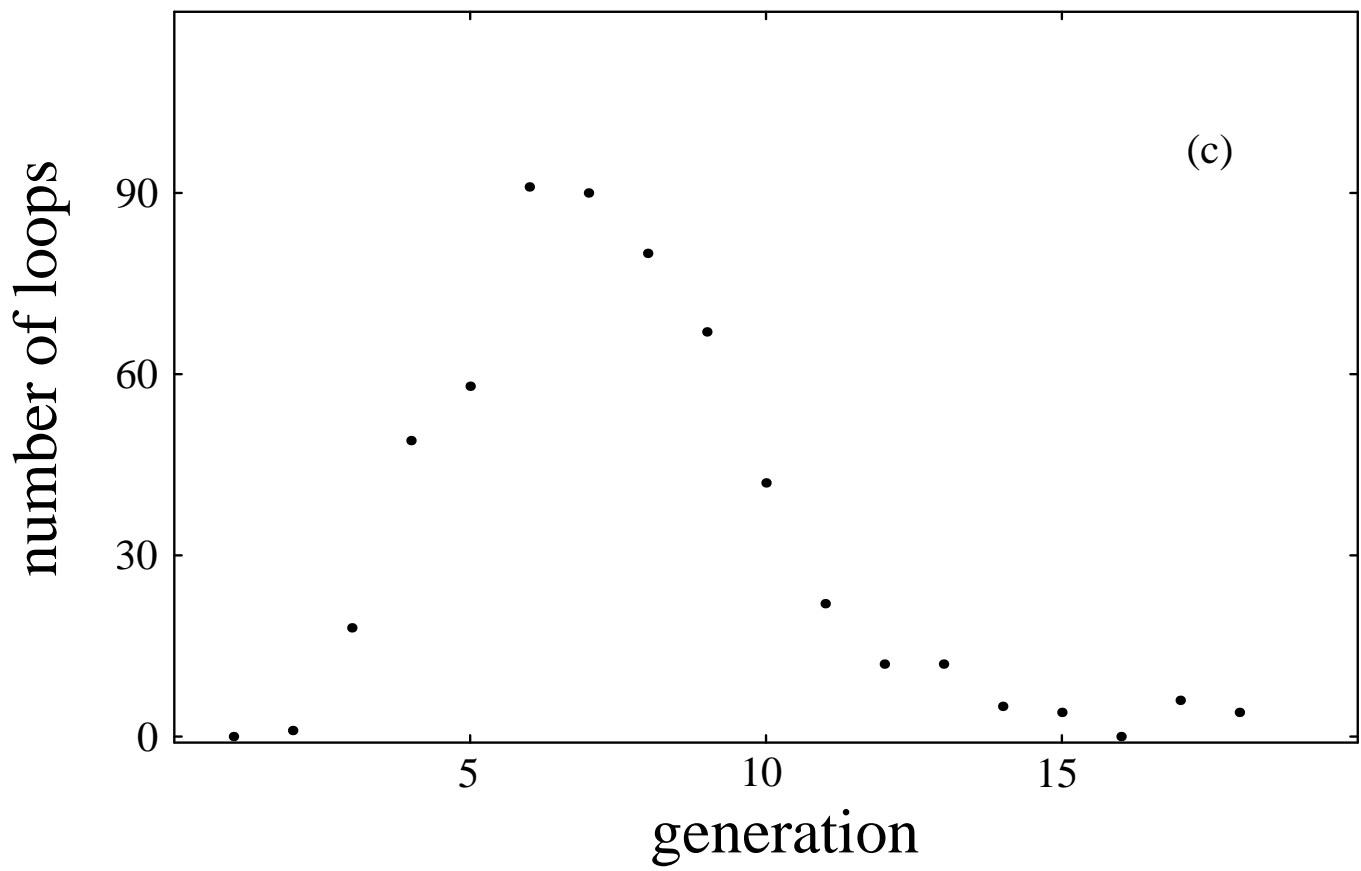

Figure 2d

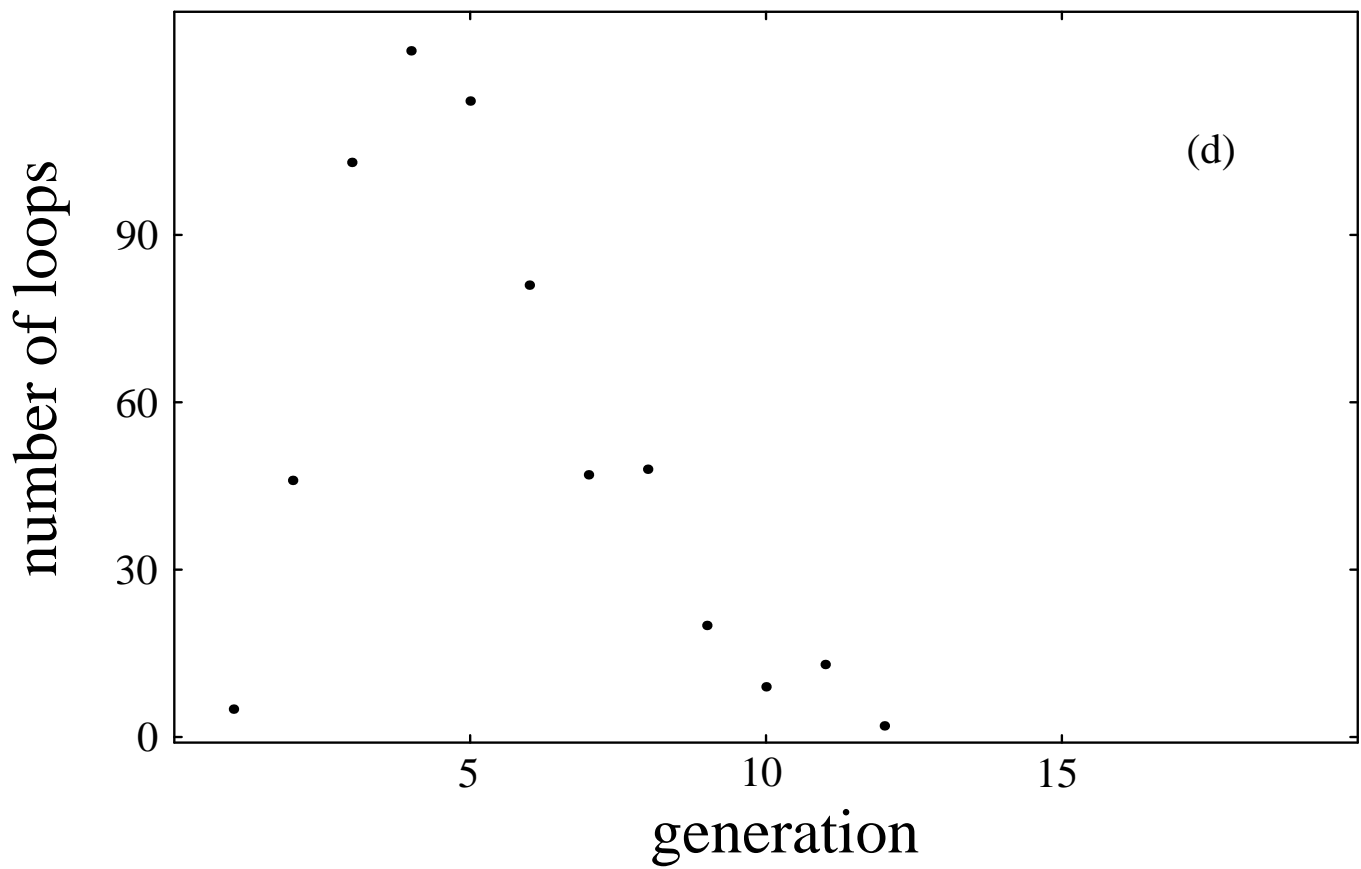

Figure 3a

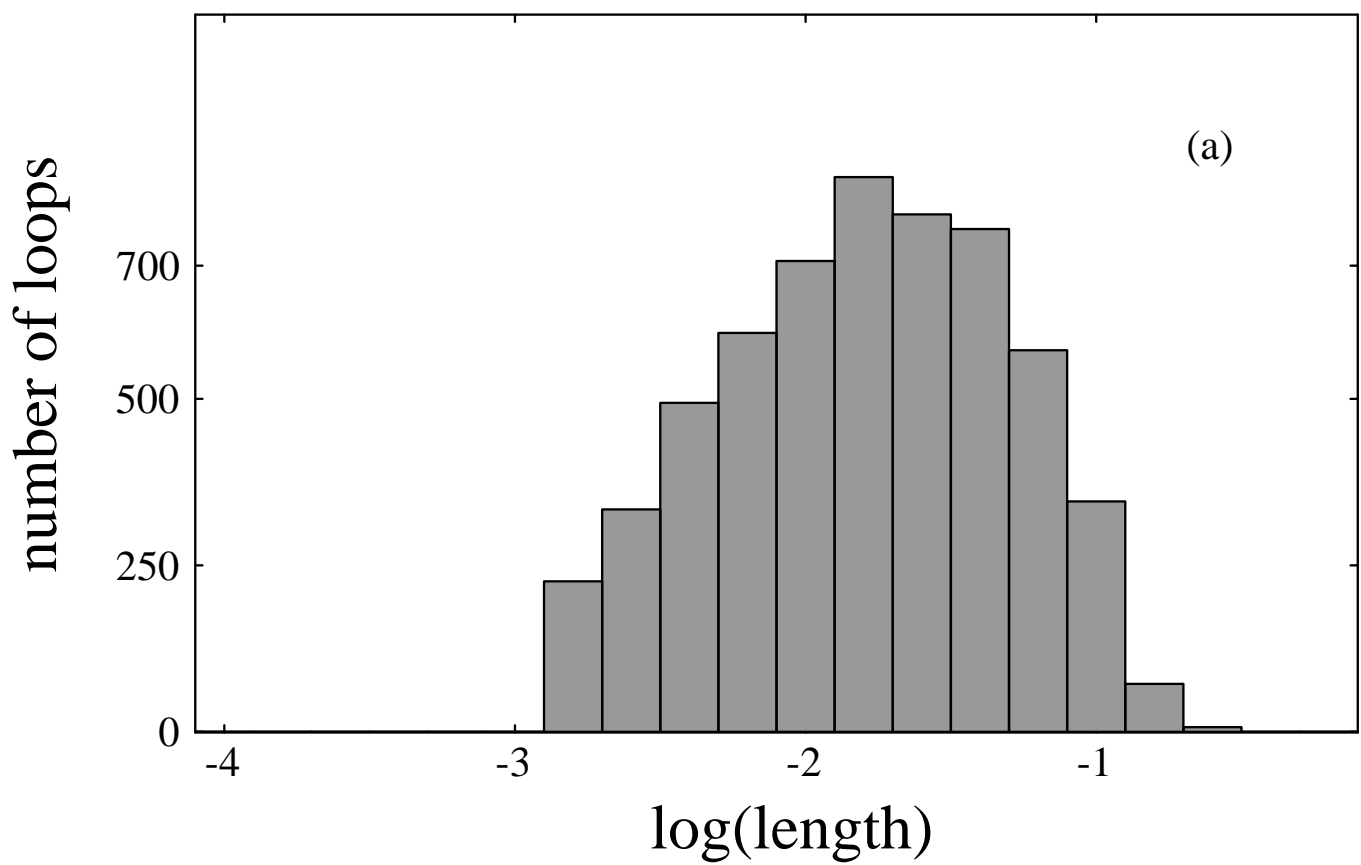

Figure 3b

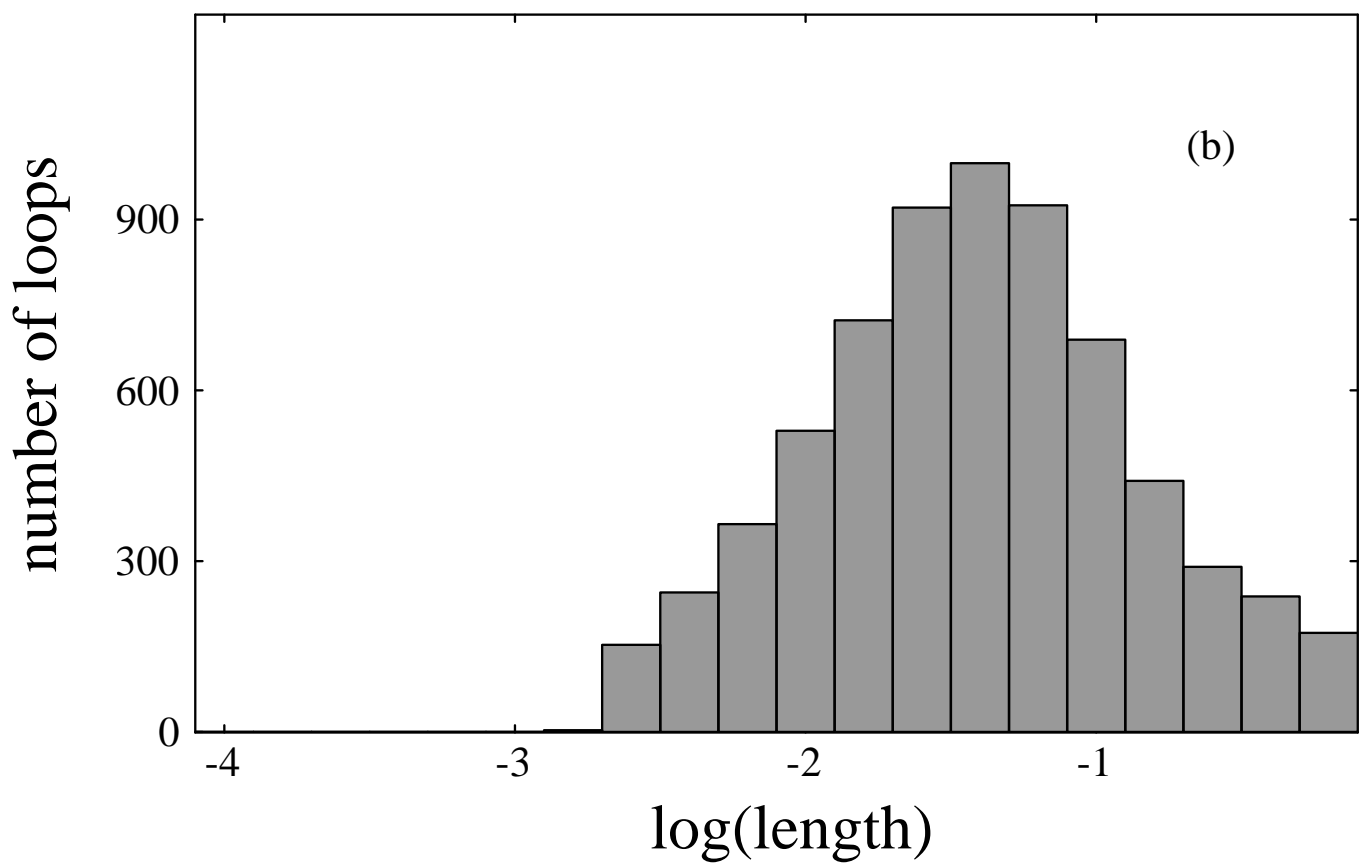

Figure 3c

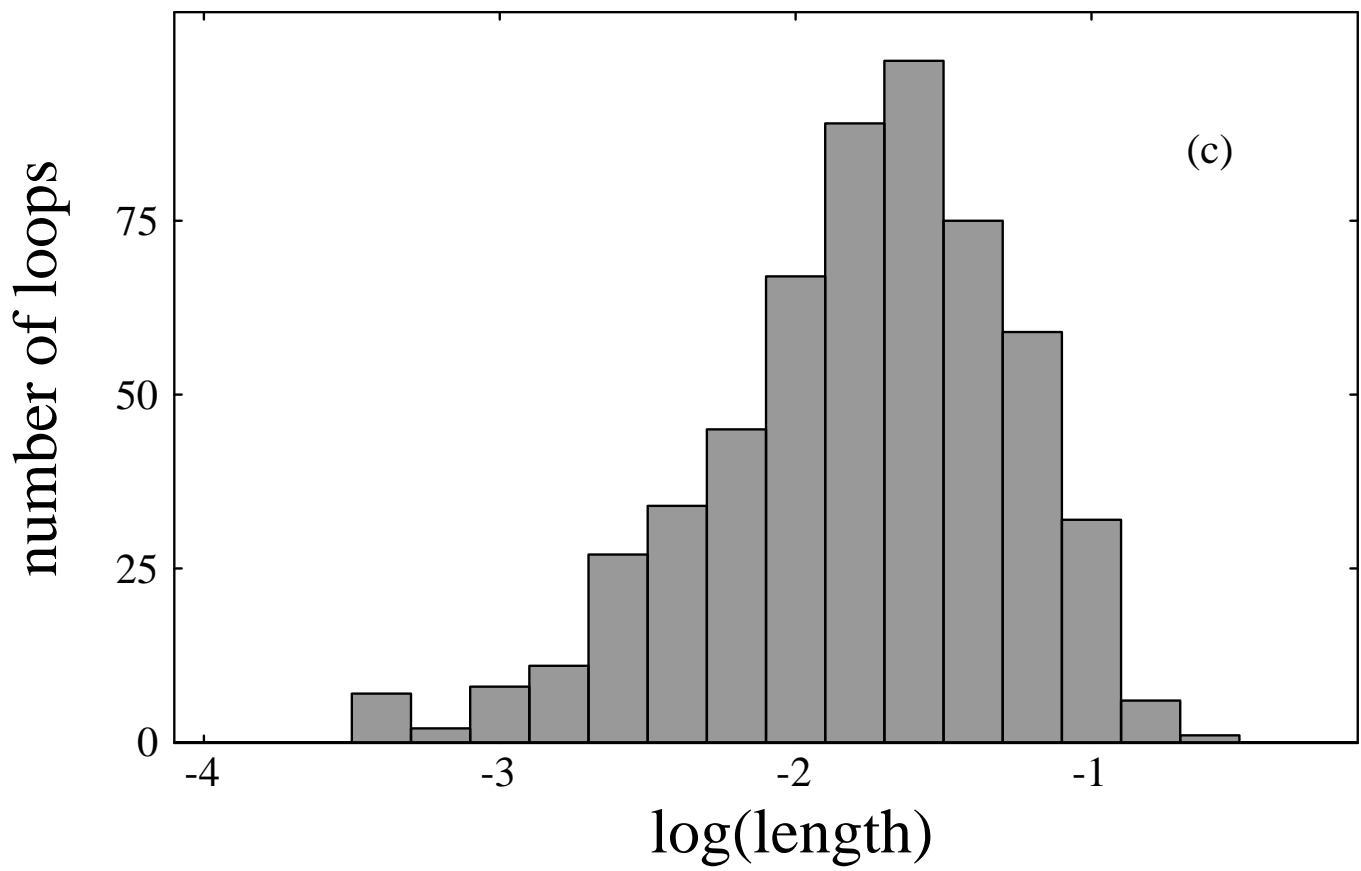

Figure 3d

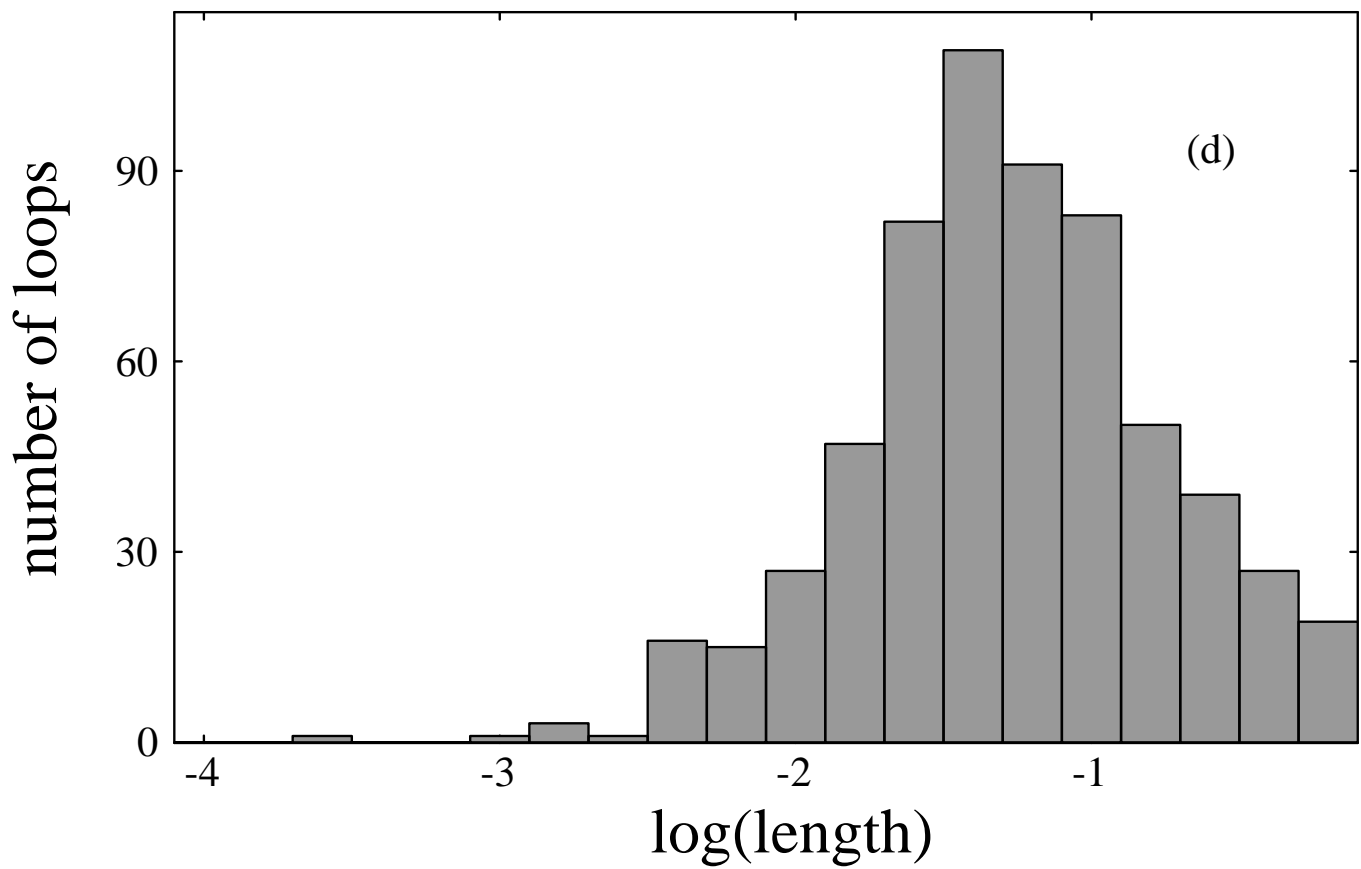

Figure 4a

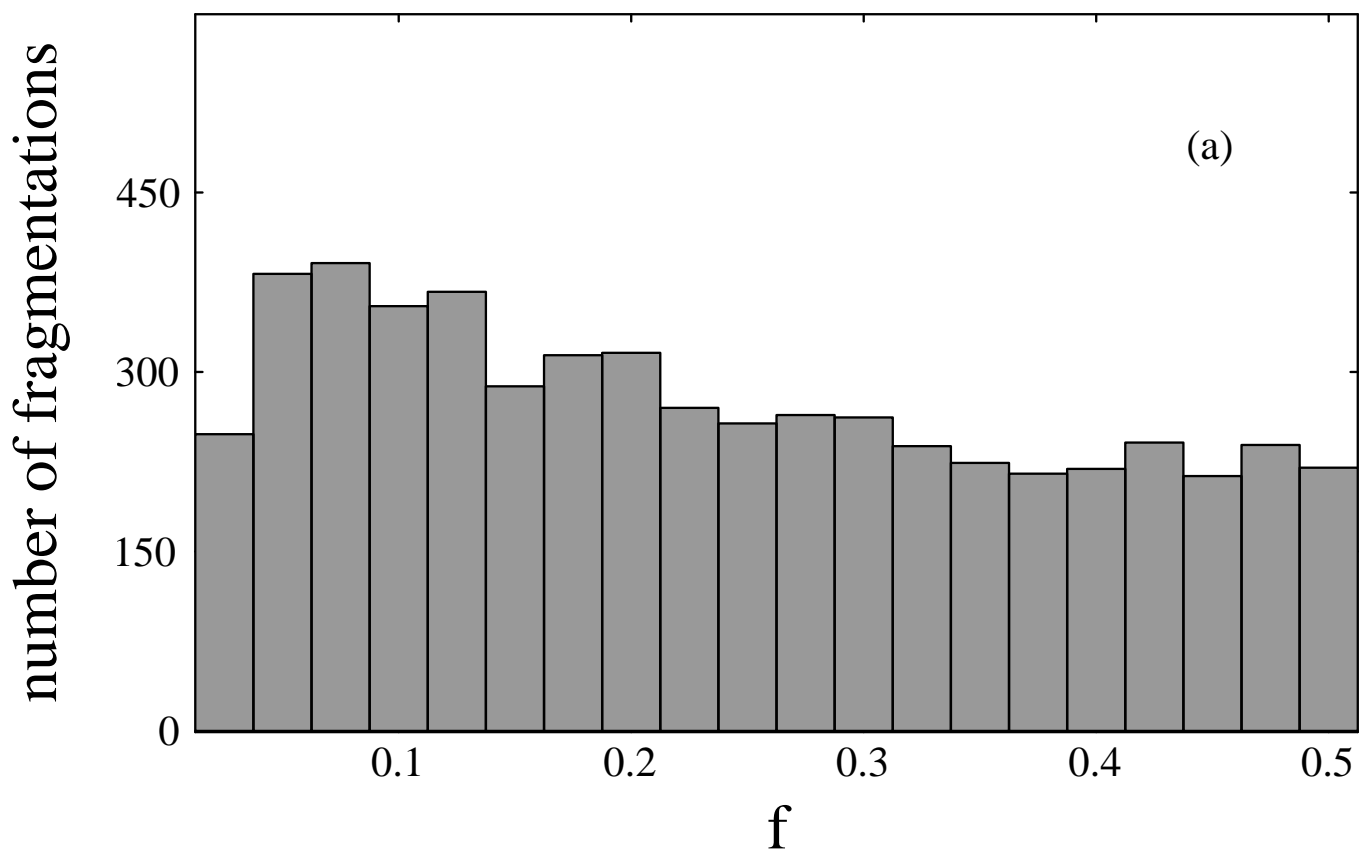

Figure 4b

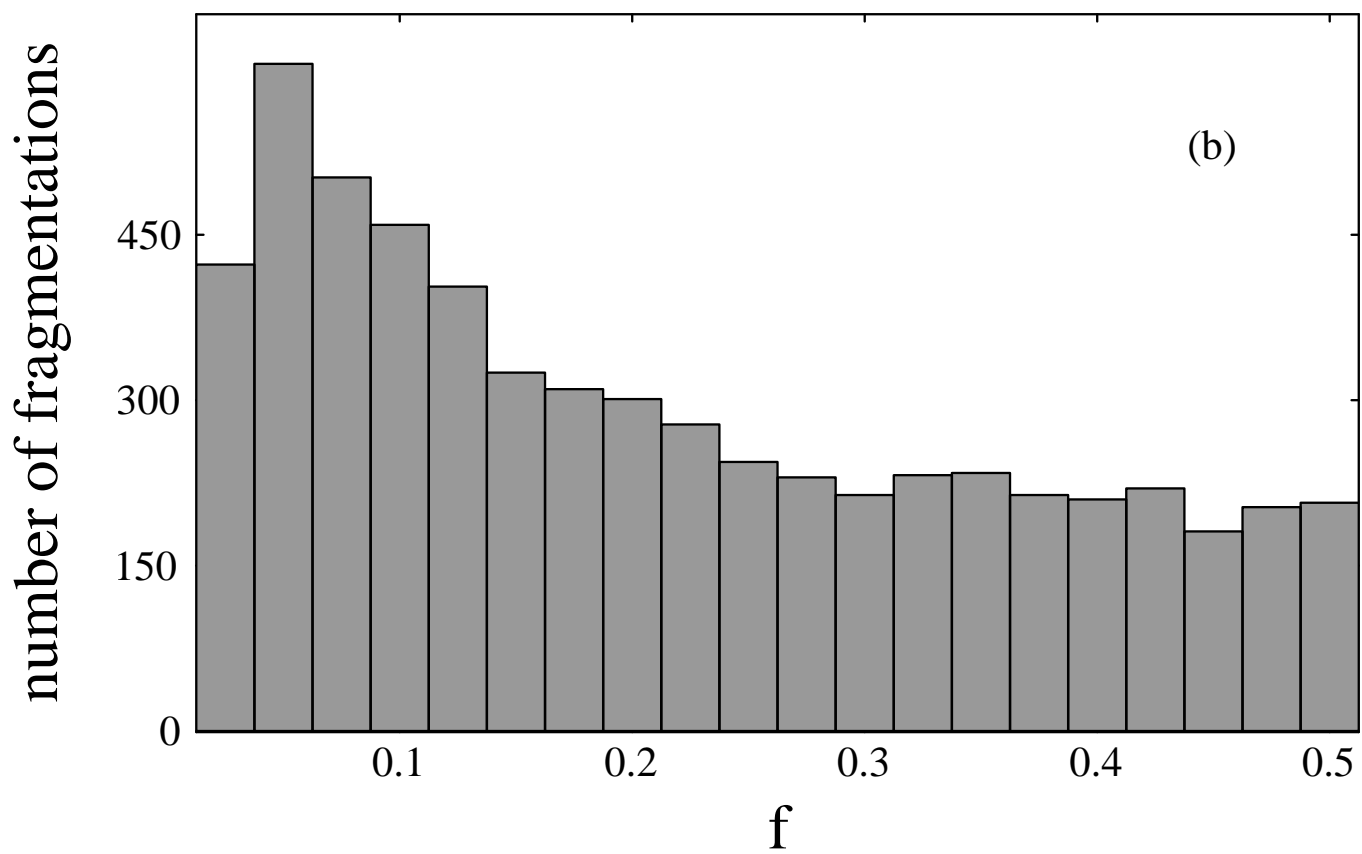

Figure 4c

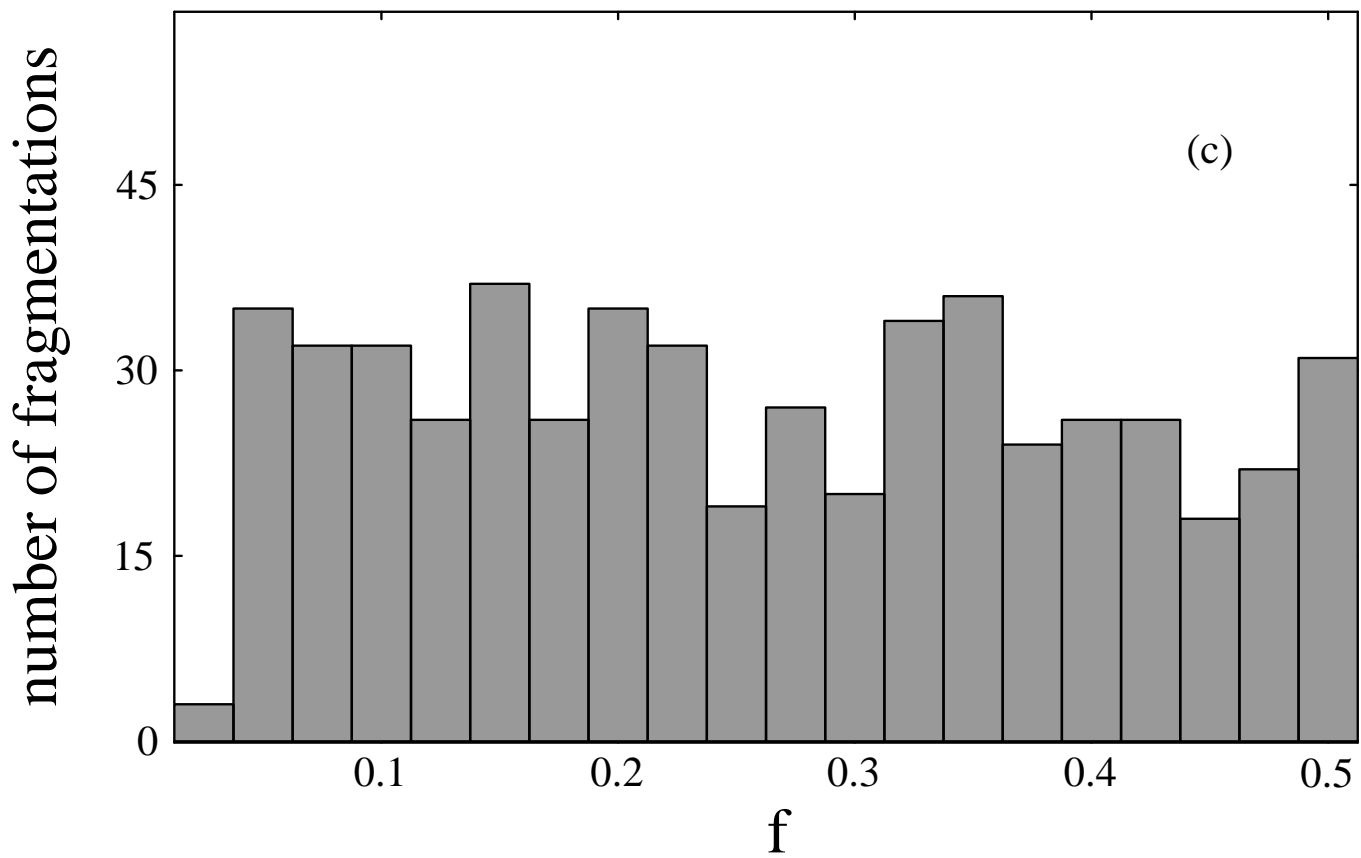

Figure 4d

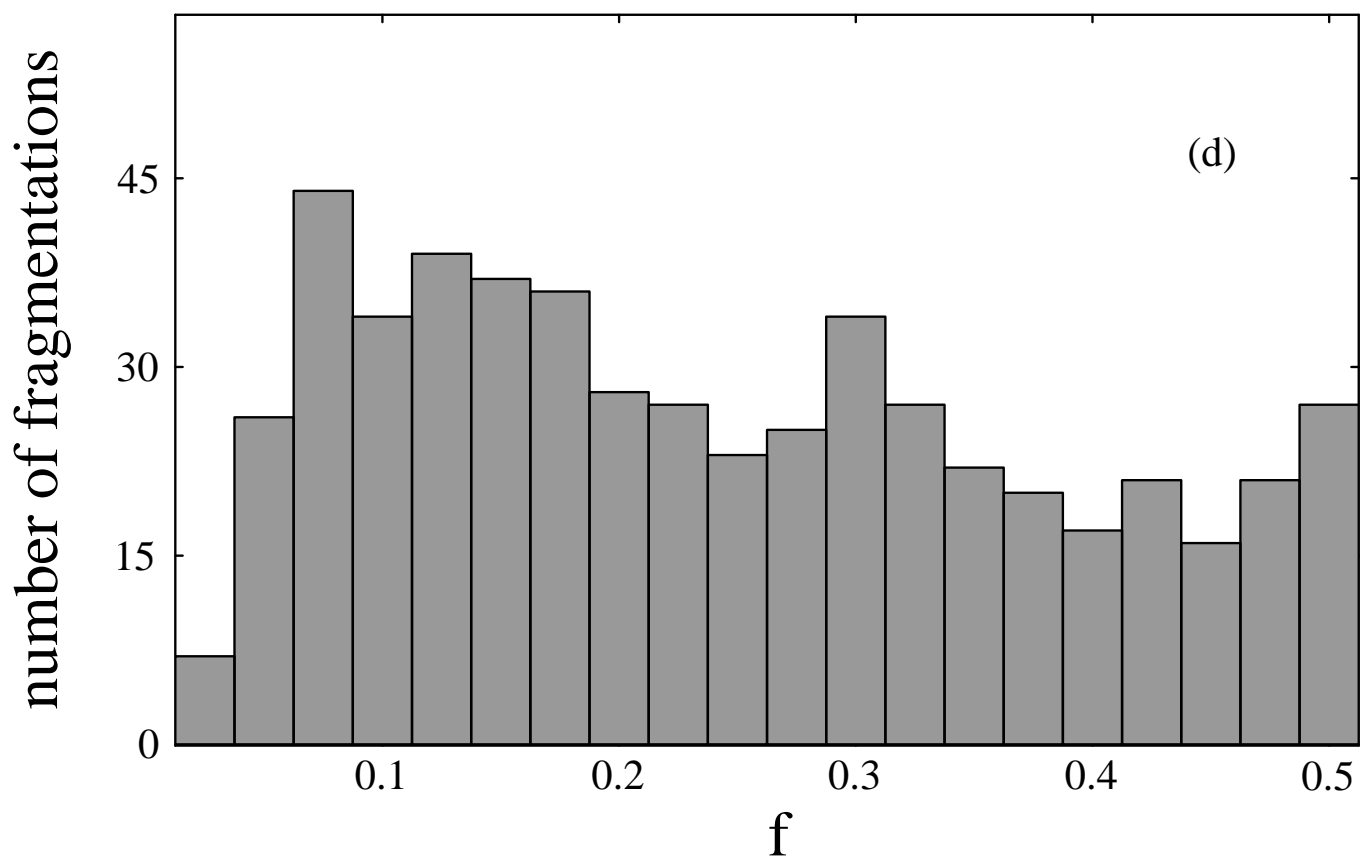

# GRAVITATIONAL RADIATION FROM REALISTIC COSMIC STRING LOOPS

Paul Casper
*Department of Physics, University of Wisconsin – Milwaukee*
*P.O. Box 413, Milwaukee, Wisconsin 53201, U.S.A.*
*email: pcasper@dirac.phys.uwm.edu*

Bruce Allen
*Department of Physics, University of Wisconsin – Milwaukee*
*P.O. Box 413, Milwaukee, Wisconsin 53201, U.S.A.*
*email: ballen@dirac.phys.uwm.edu*
(May 11, 1995)

## Abstract

We examine the rates at which energy and momentum are radiated into gravitational waves by a large set of realistic cosmic string loops. The string loops are generated by numerically evolving parent loops with different initial conditions forward in time until they self-intersect, fragmenting into two child loops. The fragmentation of the child loops is followed recursively until only non-self-intersecting loops remain. The properties of the final non-self-intersecting loops are found to be independent of the initial conditions of the parent loops. We have calculated the radiated energy and momentum for a total of 11,625 stable child loops. We find that the majority of the final loops do not radiate significant amounts of spatial momentum. The velocity gained due to the rocket effect is typically small compared to the center-of-mass velocity of the fragmented loops. The distribution of gravitational radiation rates in the center of mass frame of the loops, $\gamma^0 \equiv (G\mu^2)^{-1}\Delta E/\Delta\tau$, is strongly peaked in the range $\gamma^0 = 45 - 55$, however there are no loops found with $\gamma^0 < 40$. Because the radiated spatial momentum is small, the distribution of gravitational radiation rates appears roughly the same in any reference frame. We conjecture that in the center-of-mass frame there is a lower bound $\gamma^0_{\min} > 0$ for the radiation rate from cosmic string loops. In a second conjecture, we identify a candidate for the loop with the minimal radiation rate and suggest that $\gamma^0_{\min} \cong 39.003$.

Typeset using REVTEX

Figure 5a

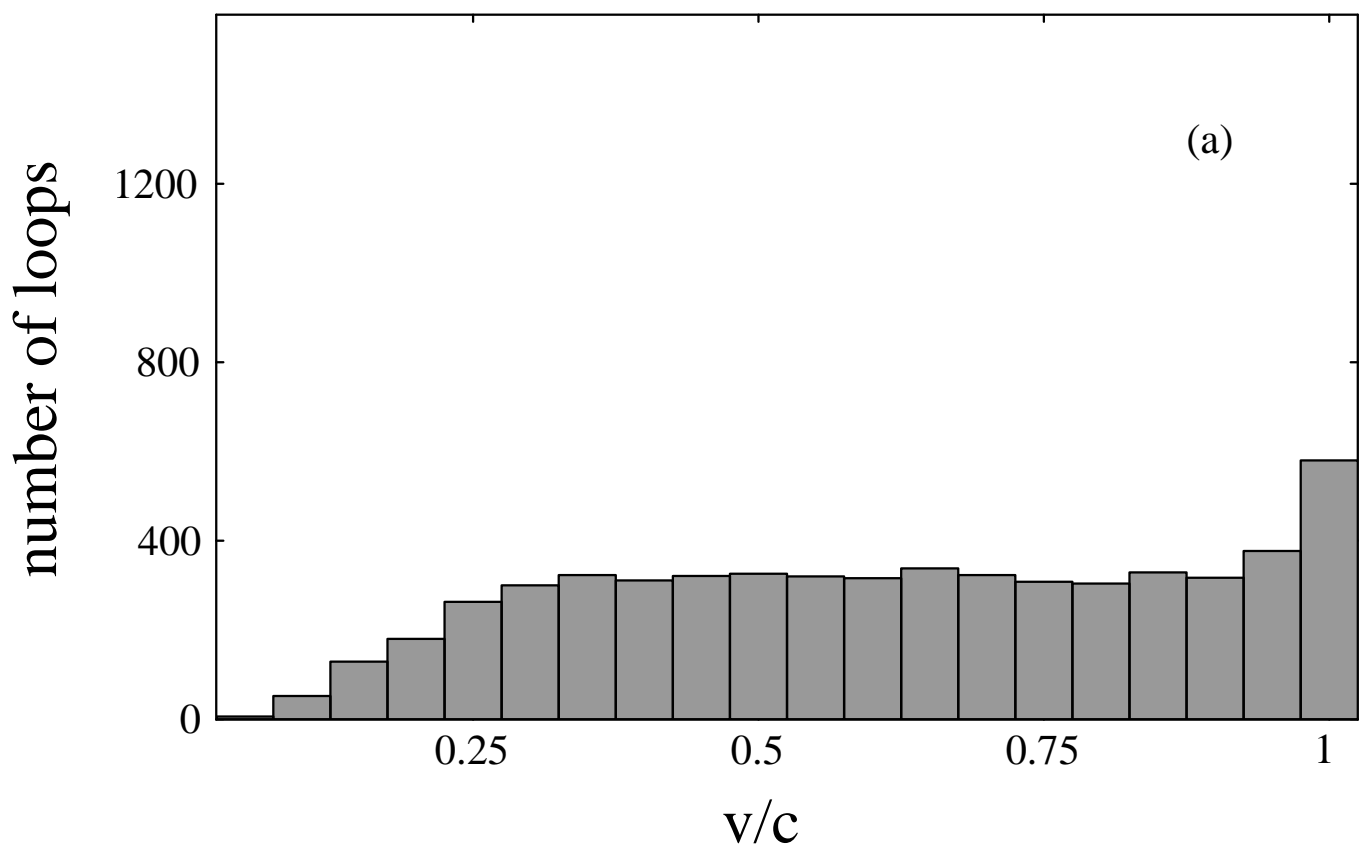

Figure 5b

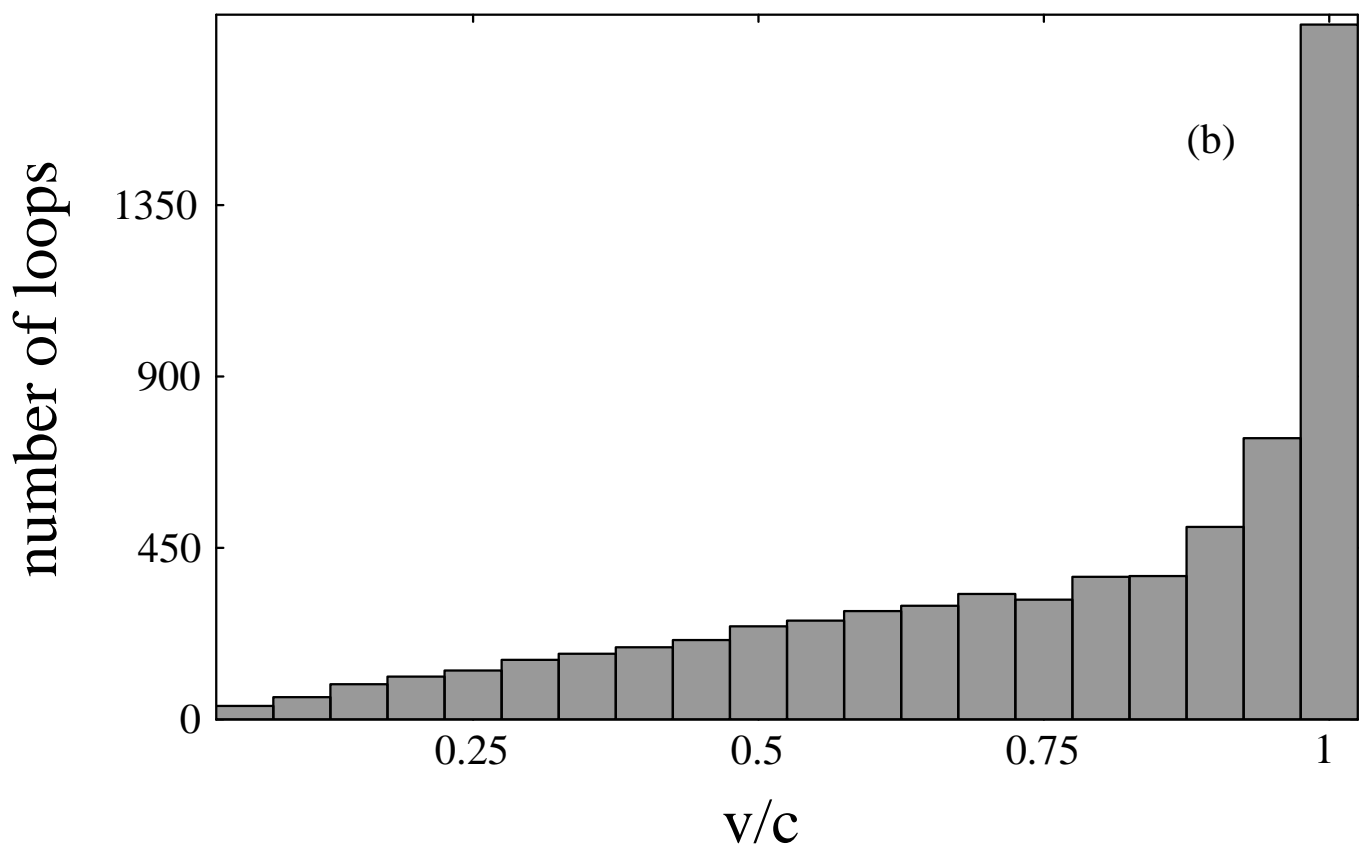

Figure 5c

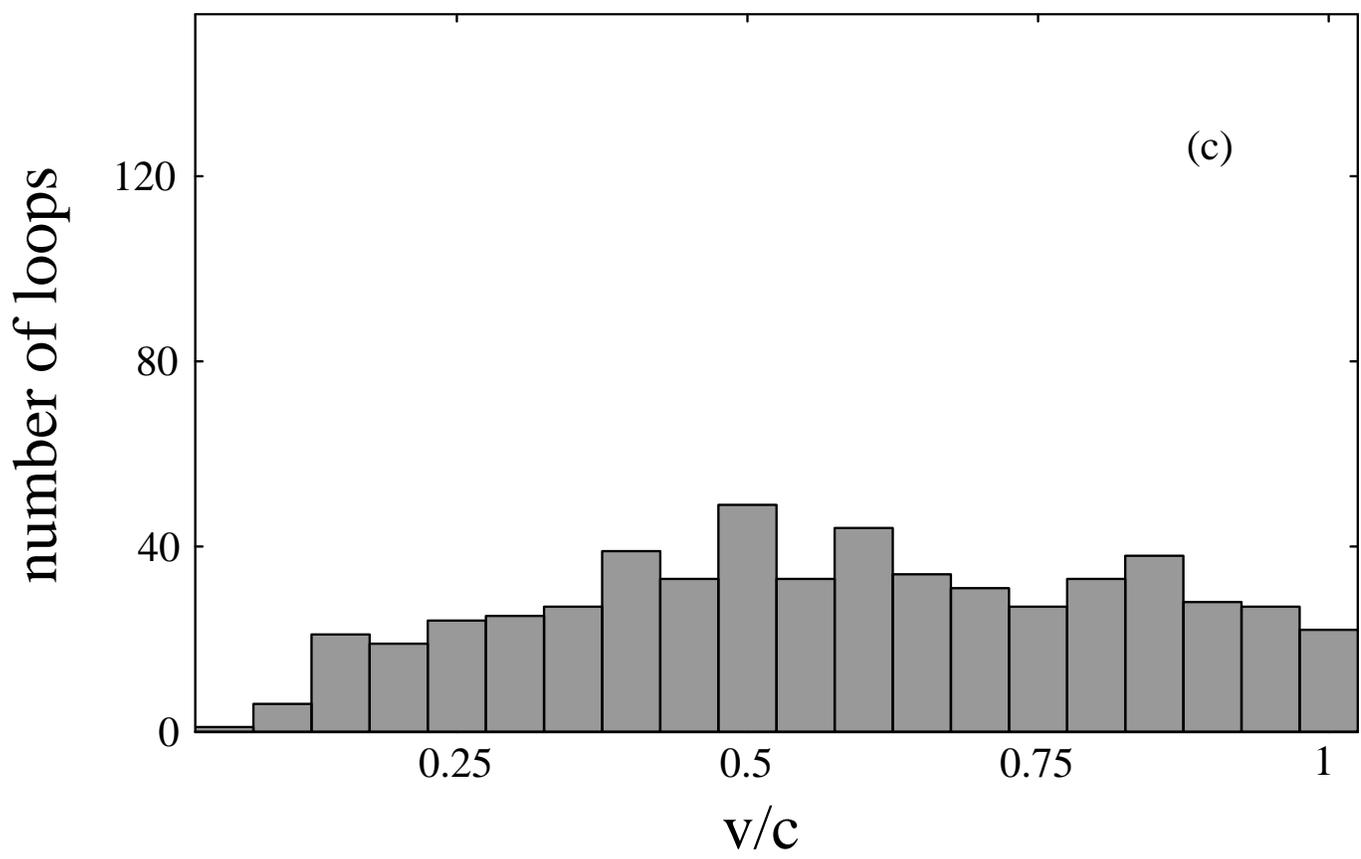

Figure 5d

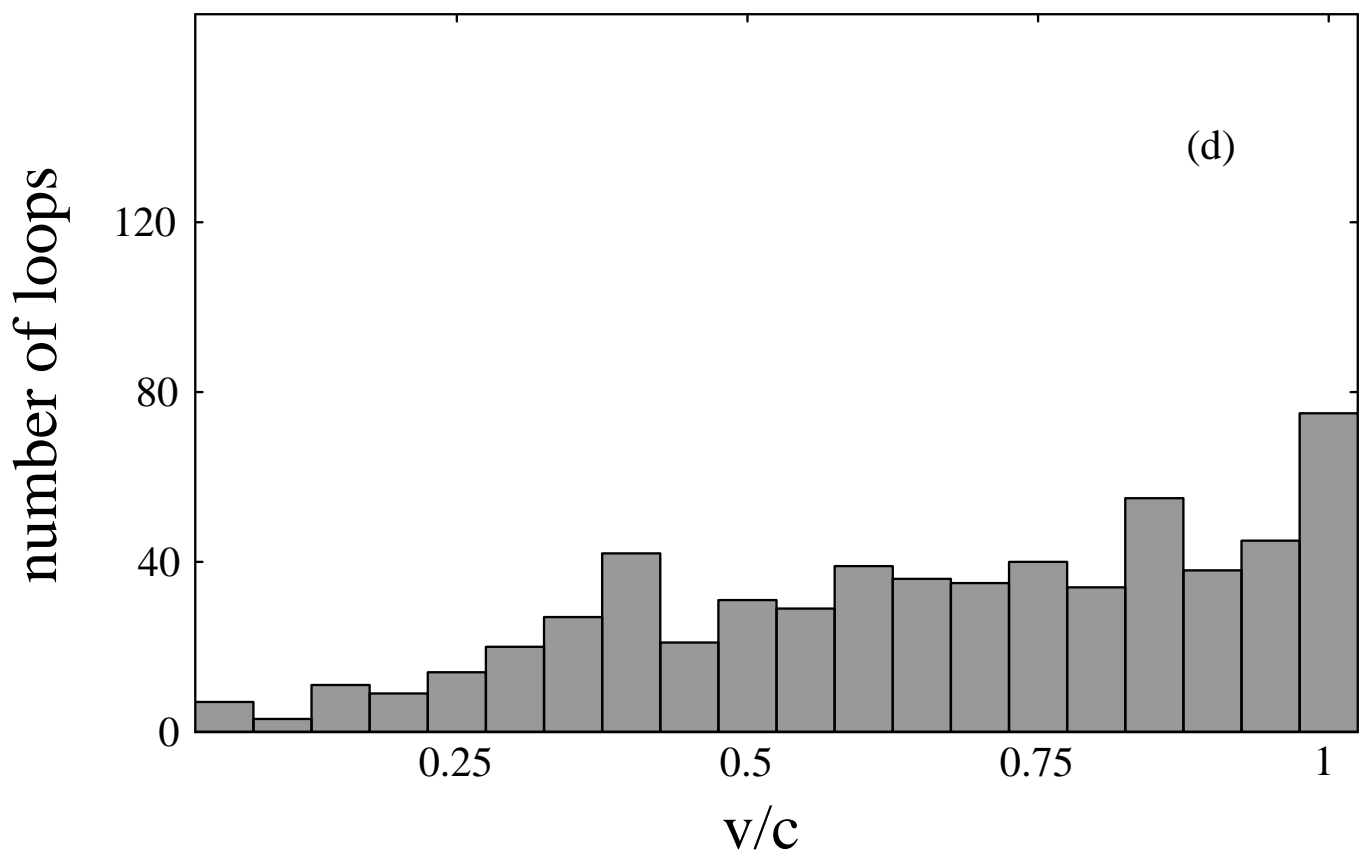

Figure 6a

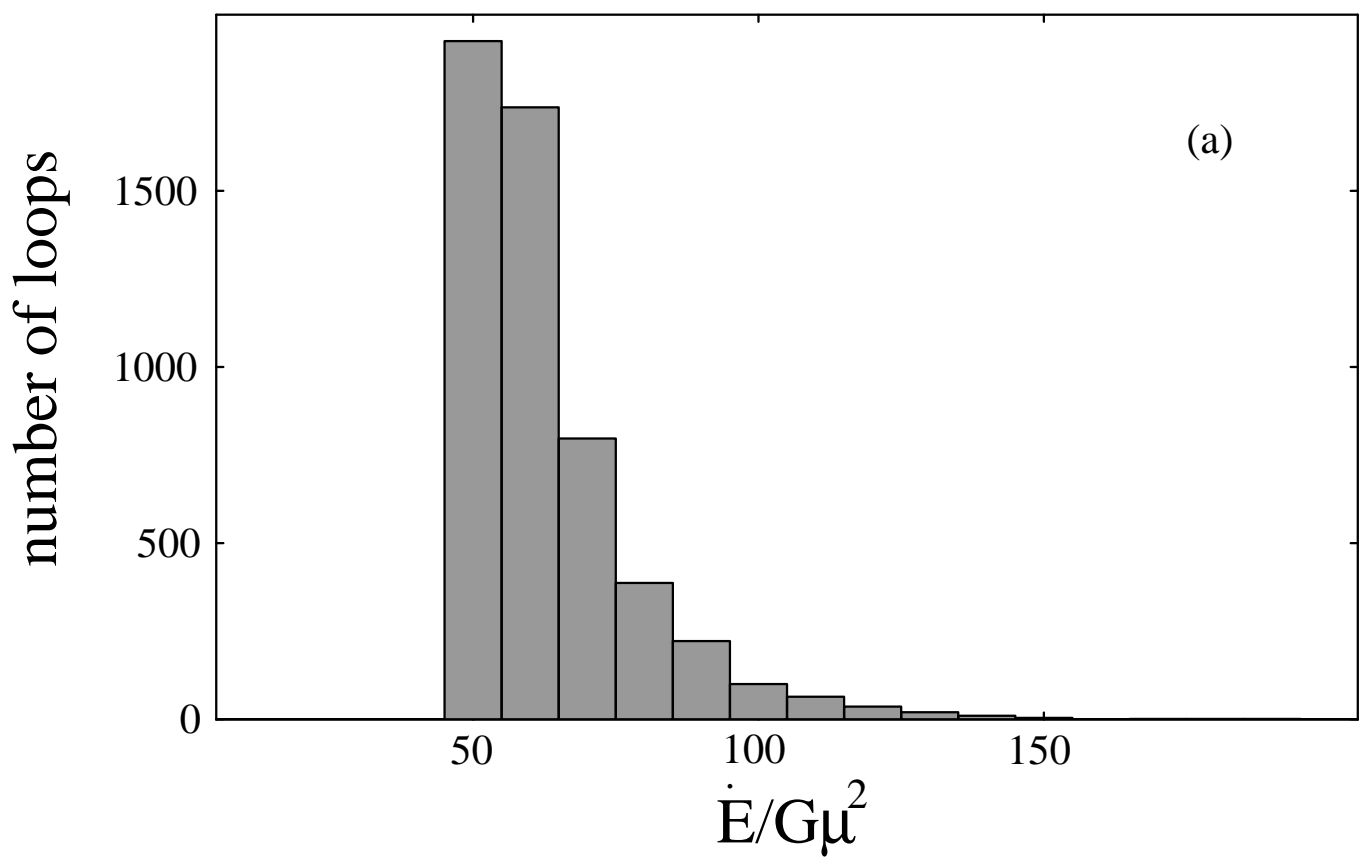

Figure 6b

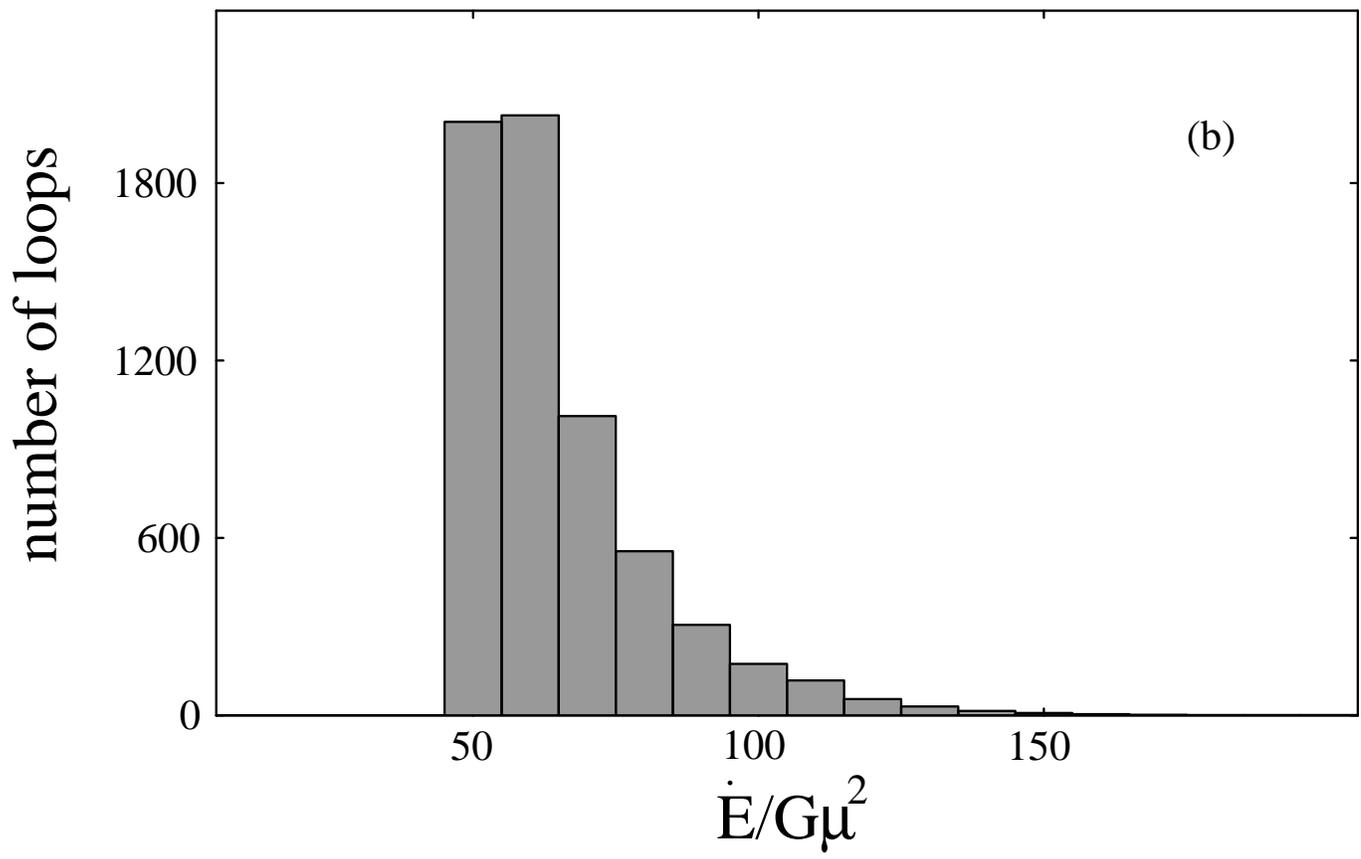

Figure 6c

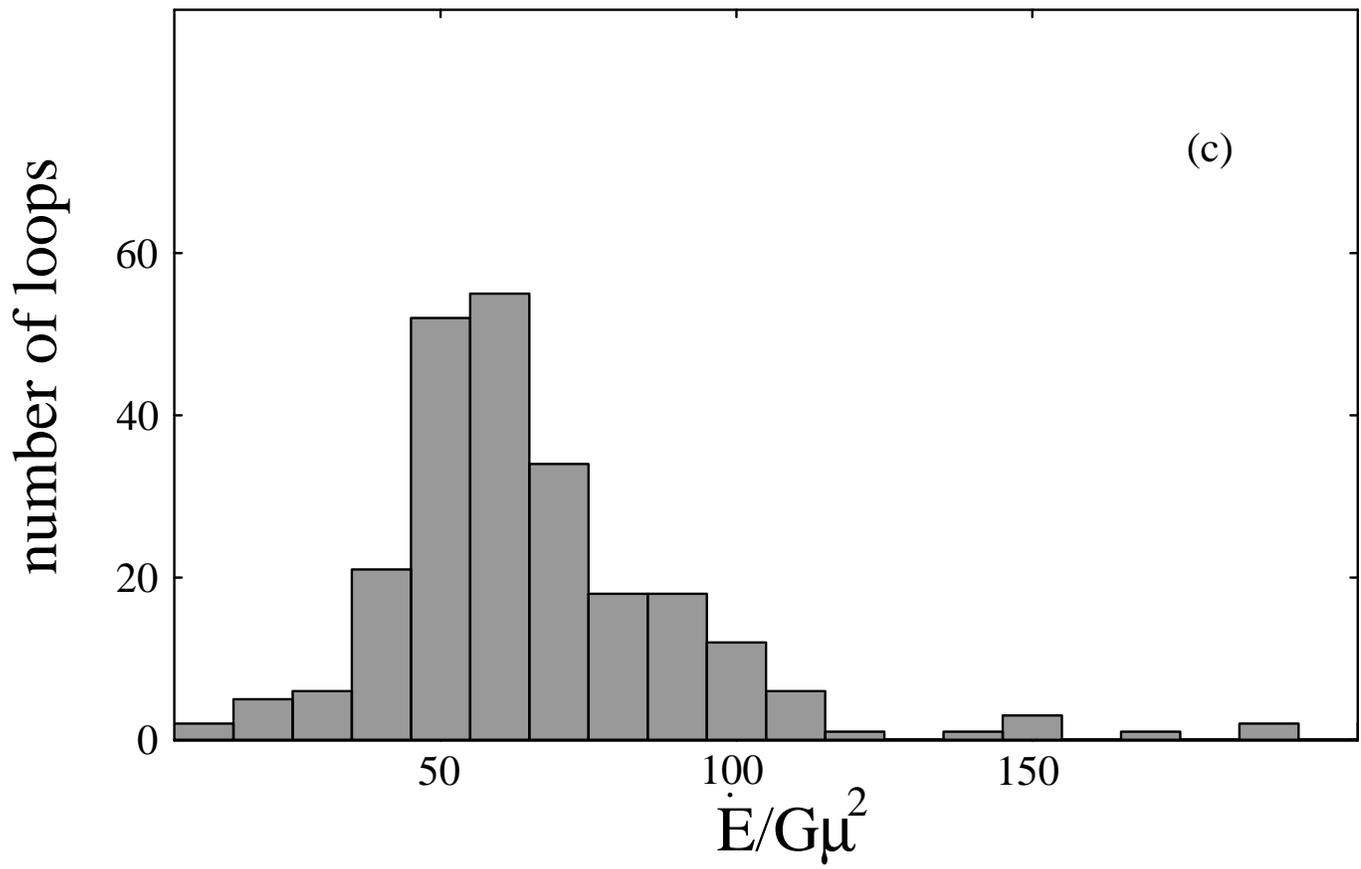

Figure 6d

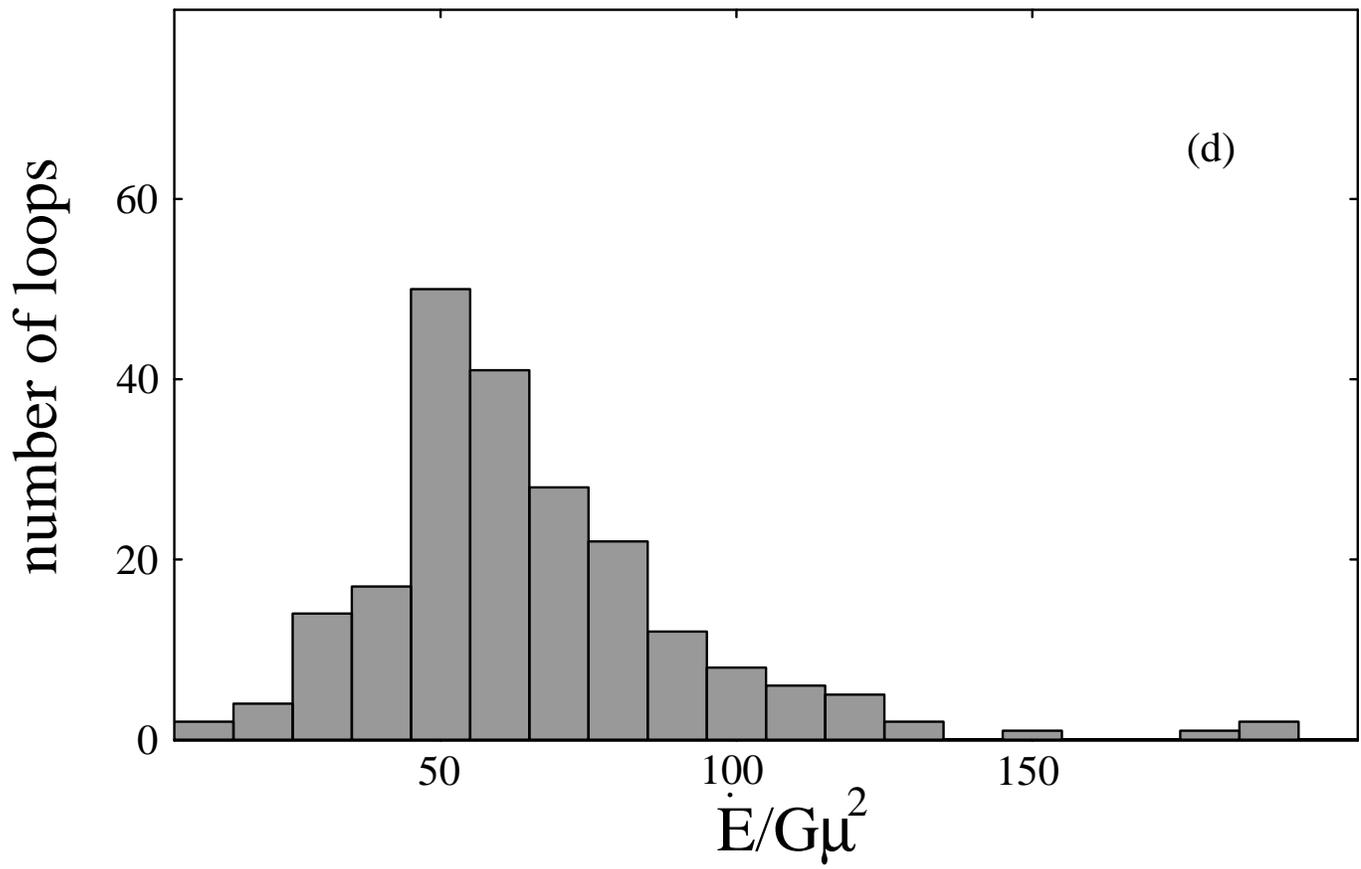

Figure 7a

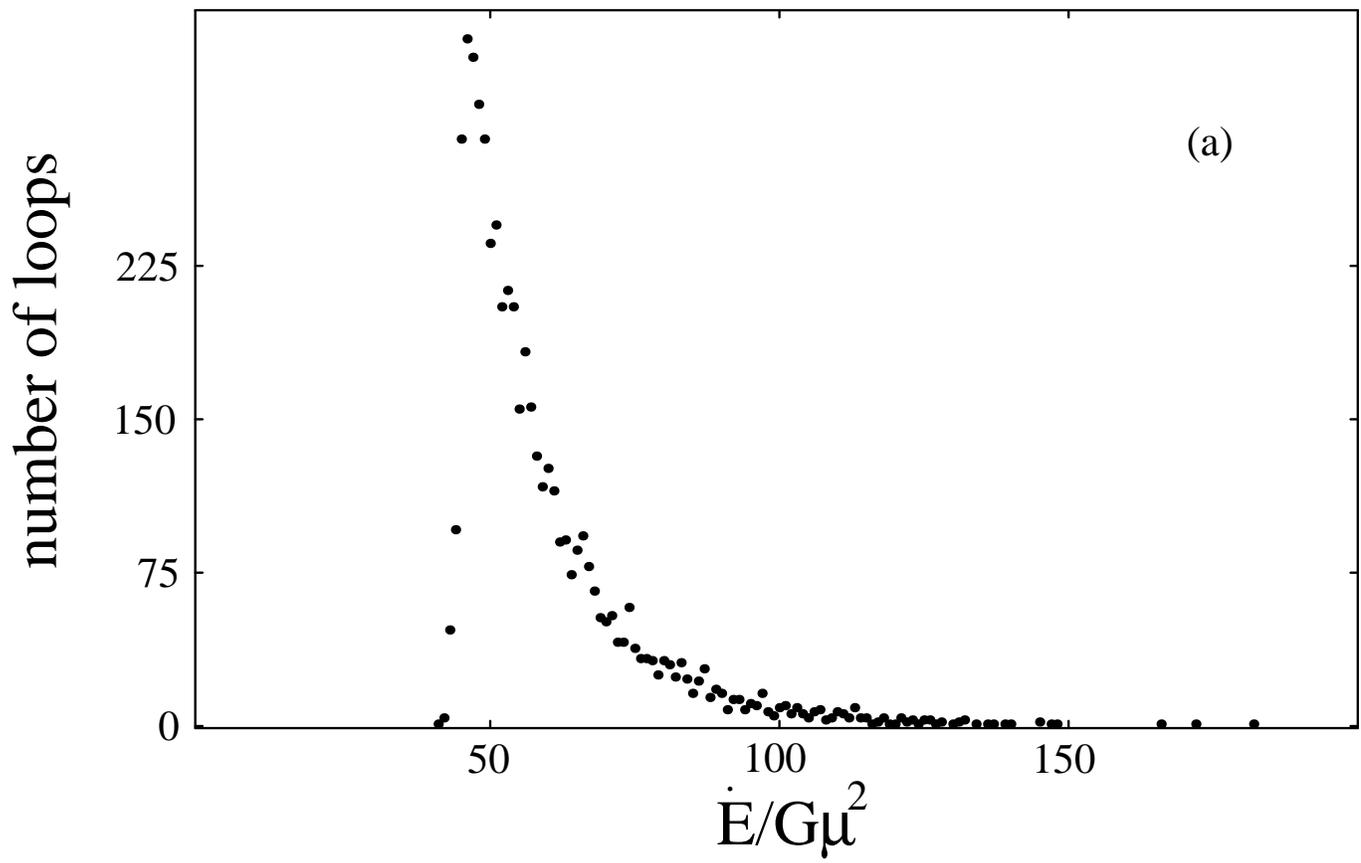

Figure 7b

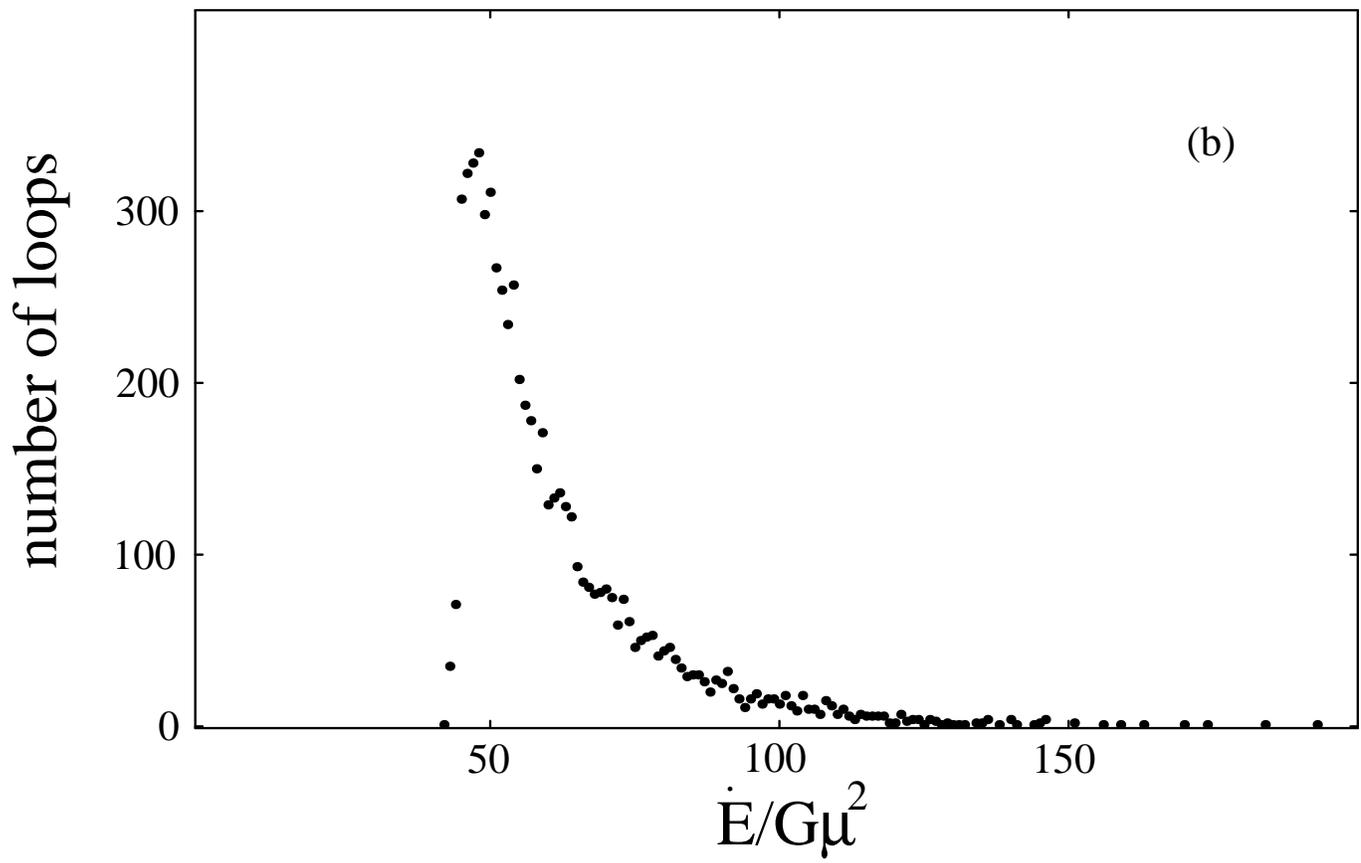

Figure 8a

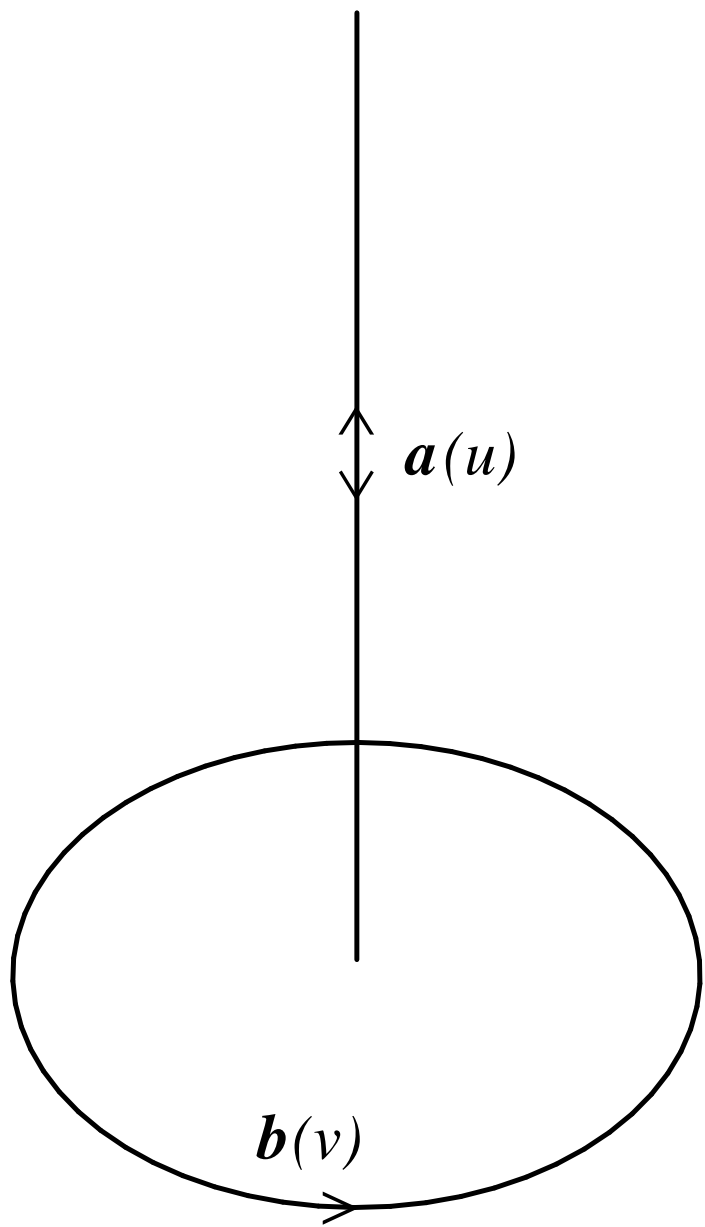

(a)

Figure 8b

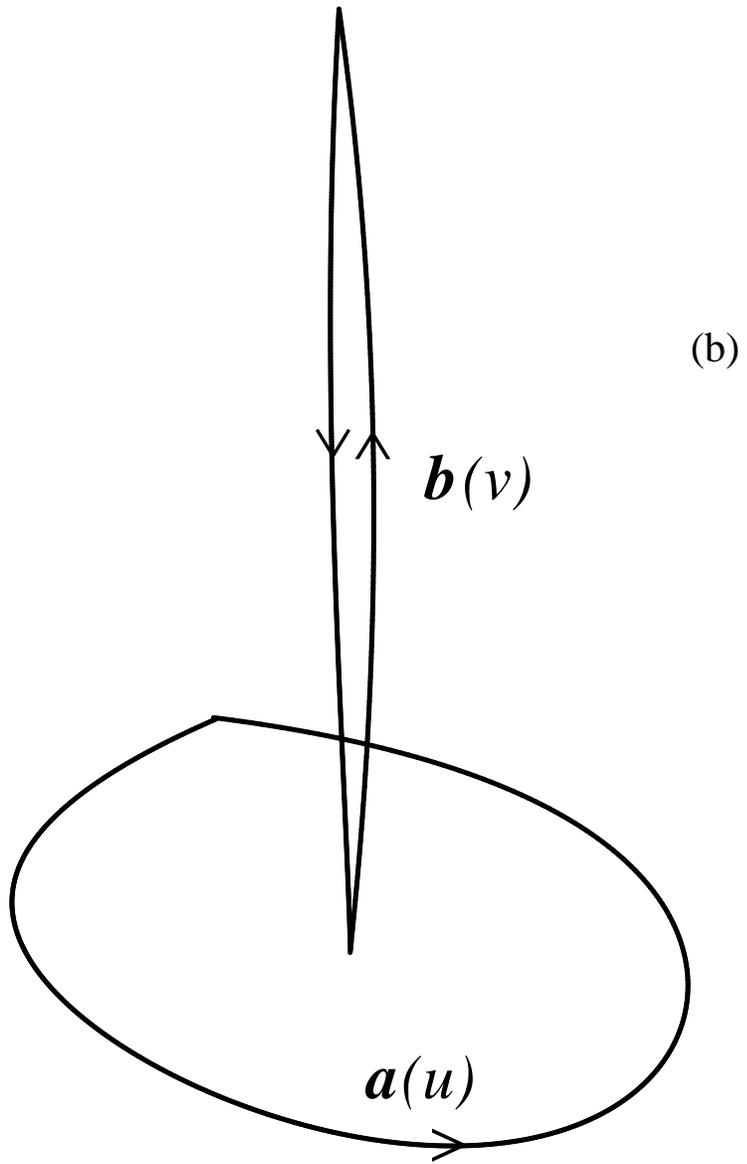

Figure 8c

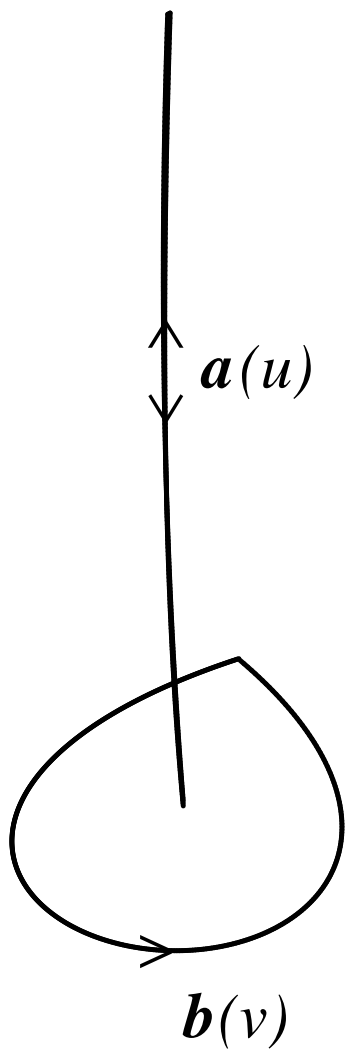

(c)

Figure 8d

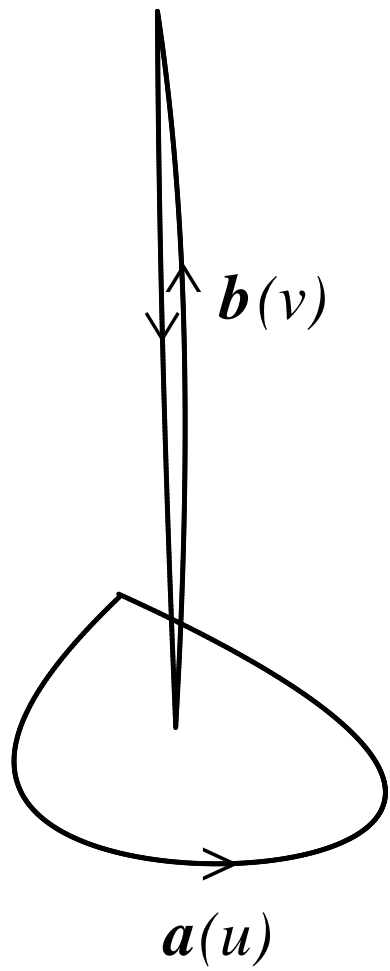

(d)

Figure 9a

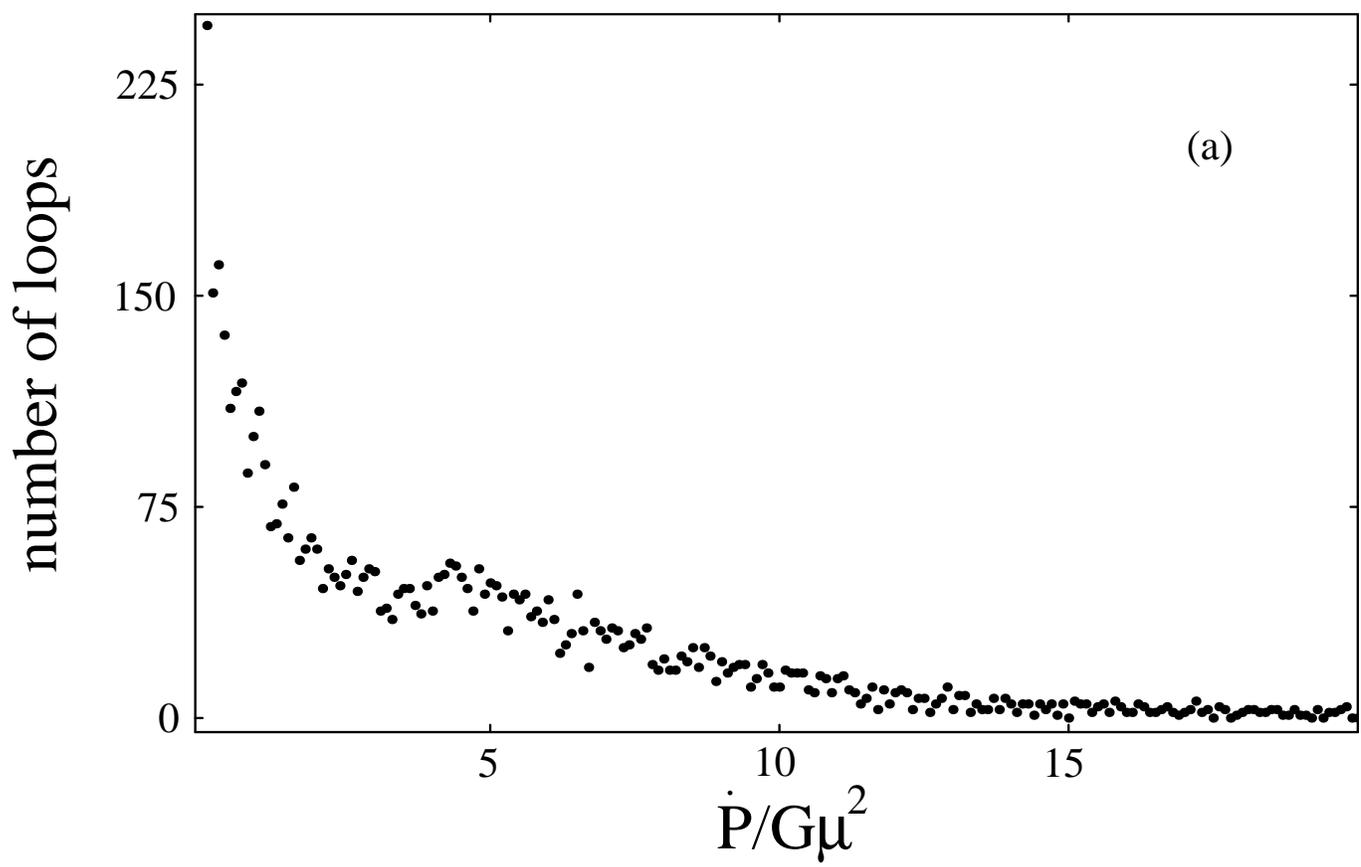

Figure 9b

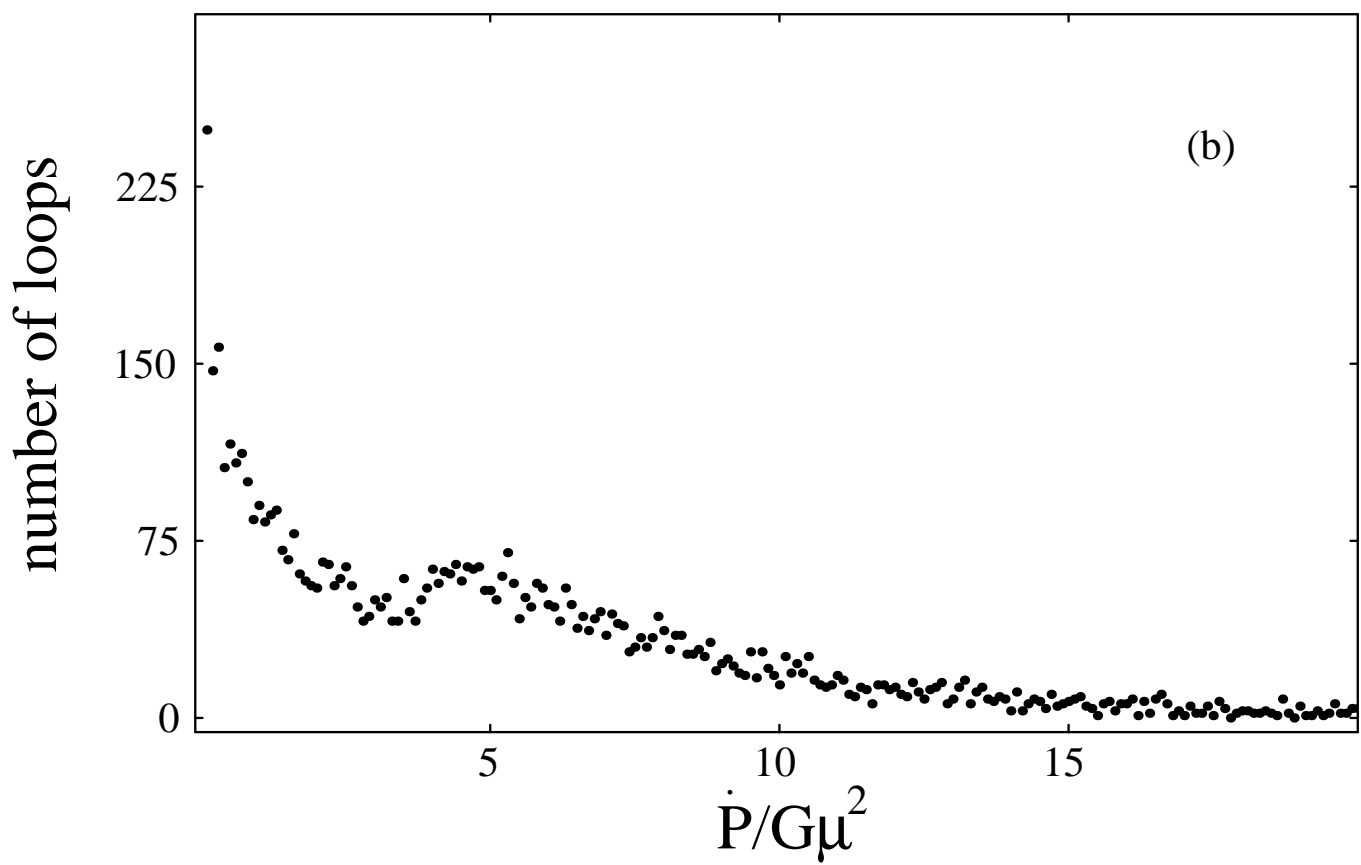